# Harnessing Large Language Model to collect and analyze Metal-organic framework property dataset

Wonseok Lee[1,†], Yeonghun Kang[1,†], Taeun Bae[1,†], Jihan Kim[1,*]

[1]Department of Chemical and Biomolecular Engineering, Korea Advanced Institute of Science and Technology, Daejeon, Republic of Korea

* Correspondence to: Jihan Kim (jihankim@kaist.ac.kr)

[†] *Wonseok Lee, Yeonghun Kang and Taeun Bae contributed equally to this paper.*

**Abstract**

This research was focused on the efficient collection of experimental Metal-Organic Framework (MOF) data from scientific literature to address the challenges of accessing hard-to-find data and improving the quality of information available for machine learning studies in materials science. Utilizing a chain of advanced Large Language Models (LLMs), we developed a systematic approach to extract and organize MOF data into a structured format. Our methodology successfully compiled information from more than 40,000 research articles, creating a comprehensive and ready-to-use dataset. The findings highlight the significant advantage of incorporating experimental data over relying solely on simulated data for enhancing the accuracy of machine learning predictions in the field of MOF research.

**Introduction**

In the past decade, there has been an increasing number of datasets available in materials science obtained from computational simulations and experimental studies.[1-6] This abundance of data is crucial for the advancement of machine learning in the field, as it underpins the development of models that can accurately predict material properties and lead to the discovery of new materials. The significance of extracting and collecting this data cannot be overstated, as it improves the accuracy of predictive models[7,8] and addresses the challenges associated with relying solely on computational simulations, which may not always reflect experimental results. This approach underscores the need to extract empirical data from the vast textual corpus of scientific literature. By mining large amounts of experimental data from already published papers, the data acquisition hurdle becomes lowered and one can enrich the connection between theoretical predictions and experimental validations in materials science.

Metal-Organic Frameworks (MOFs) are porous materials composed of metal ions or clusters coordinated with organic ligands, forming extensive networks with vast surface areas. Their significance is highlighted by a wide range of applications, from gas storage[9-13] and separation[14-17] to catalysis[17,18] and drug delivery[19,20], thanks to their customizable porosity and functionality. The sheer variety of MOFs, arising from the numerous potential metal-ligand combinations, emphasizes the critical need for comprehensive MOF data mining. This process is vital for systematically cataloging the properties and functionalities of MOFs, thus facilitating the precise design and utilization of these versatile materials across various sectors. Given the extensive potential and variability of MOFs, it is unsurprising that a substantial body of research is devoted to mining MOF data.

Previously, Park et al.[21] extracted specific data, such as surface area and pore volumes, from scientific texts using rule-based techniques, shedding light on the structural-property relationships in MOFs. Projects such as 'DigiMOF'[22] have demonstrated the effectiveness of rule-based coding in extracting synthetic information of MOFs using Natural Language Processing (NLP). The landscape has transformed with the incorporation of machine learning, as seen in the works of Nandy et al.[23,24] They leveraged predictive models to assess the stability of MOFs, marking a leap in the accuracy, and sophistication of data mining efforts. Park et al.[25] highlighted further advancements by employing Positive and Unlabeled (PU) learning to predict MOF synthesizability, showcasing the nuanced capabilities of machine learning in addressing complex challenges in MOF synthesis. Manning et al.[26] conducted a thorough analysis of synthesis protocols, focusing on ZIF-8. They identified prevailing trends and optimized methodologies, demonstrating the importance of data mining in refining synthesis processes and deepening the understanding of MOF characteristics, underscoring the significant and evolving impact of data mining techniques in the field.

Recently, the emergence of large language models (LLMs), such as ChatGPT, has revolutionized data mining, particularly in extracting nuanced information from textual sources. These models excel at understanding textual context, enabling them to perform complex data mining tasks with high efficiency. LLMs have demonstrated remarkable capability in few-shot learning, achieving accurate results with minimal examples. This reduces the threshold for adopting advanced data mining techniques across various research fields. Furthermore, prompt engineering has become crucial in optimizing LLMs for specific challenges. Yaghi's group applied LLMs to extract metal-organic framework (MOF) synthesis parameters from scientific literature with exceptional precision.[27] LLMs have a wide range of applications, including predicting crystallization outcomes and facilitating interactive platforms such as MOF chatbots, significantly enriching the paradigms of data-driven research.

Building on the pioneering work of Yaghi's group in using large language models (LLMs) for data mining, we present 'L2M3' (Large Language Model MOF Miner), an innovative data mining system that fully automates the extraction process through a sophisticated array of LLMs. Our system has rigorously analyzed over 40,000 MOF papers, extracting 32 well-defined properties in a specific format, alongside a broader collection of properties in a more general form. We have further classified the MOF synthesis process into 21 distinct categories, each with its unique data format, thereby enhancing the granularity of our dataset. Our subsequent machine learning analysis leverages this rich dataset to elucidate the critical role of experimental data in advancing MOF research, demonstrating the profound impact that systematic data collection has on the field.

# Result & Discussion

*Workflow*

a)

Figure: Workflow diagram showing Data Extraction Agent (processing Tables and Texts from Literature papers into Synthesis conditions and Characteristic properties) and Data Organizing Agent (organizing Table data, Synthesis data, and Property data, with Matching against CCDC database into L2M3 Database).

b)

Figure: Table processing pipeline — Markdown table → Categorization (Category: Property) → Inclusion (List of Properties: [LCD, PLD, $T_d$]) → Extraction (JSON output with 'meta': {'name': '1(a)'}, 'LCD': {'value': 5.5, 'unit': 'Å'}, 'PLD': {'value': 4.3, 'unit': 'Å'}, '$T_d$': {'value': 471, 'unit': 'Å'}, {'meta': {'name': '2(b)'}, 'LCD': {'value': 7.1, ....)

c)

Figure: Text processing pipeline — Texts ("The unique structure of MOFZ contributes to its remarkable pore volume, measured at 0.8 cm³/g. This high pore volume, crucial for accommodating gas molecules, positions MOFZ as an efficient adsorbent.") → Categorization (Property: O, Synthesis: X) → Inclusion (List of Properties: [Pore Volume]) → Extraction (JSON: { 'meta': {'name': 'MOFZ'}, 'pore volume': {'value': 0.8, 'unit': 'cm³/g'} })

Figure 1. (a) Overall schematic of the L2M3 model (b) Overall process of table mining (c) Overall process of text mining

Figure 1(a) shows the overall framework of the L2M3 model, which extracts information from both text and tables in scientific papers. The model employs three specialized agents: the table agent, the synthesis condition agent and the property agent, all of which are used in the mining process. These three agents are used to extract information from papers about the properties and synthesis conditions of MOFs. After extraction, a matching agent standardizes material names and symbols to consolidate the extracted data into a unified metadata set. Finally, the consolidated dataset is matched against crystal data in the CCDC database[28] for structural matching, as shown in Supplementary Figure S1. The process is designed for efficient input and output handling through a single code execution.

Figure 1(b) explains how the table agent processes table data. Achieving accurate extraction from complex tables using rule-based coding techniques is challenging. To address this, a table agent based on LLM has been developed to increase the reliability of information extraction from tables. This agent works in three stages: categorization, inclusion and extraction. First a categorization agent sorts tables into different types. Examples of each table type are provided in Supplementary Table S1-3. Next, inclusion is performed on the property table and the crystal information table, determining what information is contained in these tables. Finally, the extraction agent retrieves relevant information from both the Property and Crystal Info tables.

Figure 1(c) shows the process used by the synthesis condition agent and the property agent to extract information from text. In categorization step, the agent classifies the texts based on whether they describe a property, a synthesis condition, or contain no relevant information. If a paragraph describes a synthesis condition, the synthesis condition agent is used; if it refers to a property, the property agent is used. Each agent then performs inclusion steps to determine the specific information present. The properties and synthesis methods that can be extracted are listed in Supplementary Table S4-5. Following the inclusion steps, the extraction steps are performed to extract the information. The extraction prompt is tailored to the synthesis condition or property type identified, increasing accuracy and correctly formatting the data.

*Data mining result*

After analyzing more than 40,000 academic papers and excluding those with errors, we successfully collected data from 39,476 papers related to synthesis conditions and properties. To evaluate the accuracy of our data analysis, which was performed using LLM, we randomly selected 150 papers. The evaluation was conducted separately for the type of task and the type of paragraph described. Additional information on our grading process and other important factors can be found in Supplementary Note S1.

| Data Type | Categorization | | | Inclusion | | | Extraction | | |
|---|---|---|---|---|---|---|---|---|---|
| | Precision | Recall | F1 Score | Precision | Recall | F1 Score | Precision | Recall | F1 Score |
| Synthesis | 1.00 | 0.98 | 0.99 | 0.96 | 0.91 | 0.94 | 0.96 | 0.90 | 0.93 |
| Property | 0.98 | 0.95 | 0.97 | 0.98 | 0.98 | 0.98 | 0.97 | 0.90 | 0.93 |
| Table | 0.99 | 1.00 | 1.00 | 1.00 | 1.00 | 1.00 | 1.00 | 1.00 | 1.00 |

Table 1. Precision, recall, and F1 score of each step

In the categorization task, our research achieved F1 scores higher than 0.95 across all three categories, outperforming previous studies in terms of accuracy. Similarly, the inclusion task consistently scored F1 scores of 0.95 or higher in every instance. Lastly, our extraction task demonstrated enhanced accuracy compared to past efforts in most situations. A detailed comparison between other papers and this work can be found in the Supplementary Note S2.

As noted in Supplementary Note S2, previous studies on text mining for extracting properties MOFs have been limited to only one or two specific properties. In contrast, our research is able to extract a wide range of properties, encompassing over 20 different attributes, and surpasses the accuracy achieved in previous works. Furthermore, our data mining tool demonstrates higher accuracy in identifying synthesis conditions compared to previous efforts. Moreover, our tool covers not only the synthesis condition but also the entire synthesis process, including pre-processing and post-processing steps, unlike earlier research that mainly focused on the synthesis condition.

As the number of papers increases, the diversity of paper formats also increases. This can lead to a decrease in the accuracy of mining. Previous studies have shown a trade-off between accuracy and the number of papers extracted, as shown in Supplementary Note S2. Some studies have high accuracy but extract only a few papers, while others extract a large number of papers but with lower accuracy. However, this study achieved high accuracy despite extracting from a large number of papers.

*Statistics*

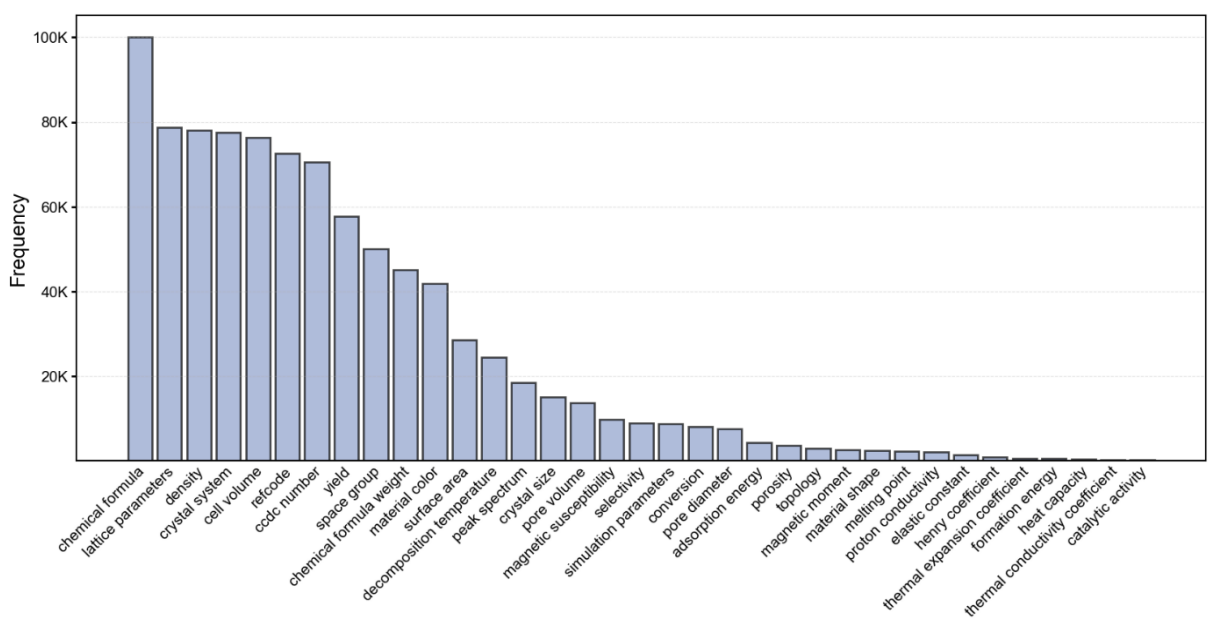

Figure 2. Distribution of mined properties of Metal-organic framework

Through the L2M3 system, we have collected information on synthesis conditions and various physical and chemical properties of the MOF material group. Figures 2~3 depict the distribution of these properties in graphical form, and statistics related to synthesis conditions can also be found in Supplementary Figure S2~3. Figure 2 shows the distribution of properties which we have clearly defined formats in JSON, revealing that information on chemical formula, lattice parameters, and density, which are necessary to estimate in the synthesis of MOF materials, are most frequently reported. We have also gathered a substantial amount of data on properties that are difficult or impossible to obtain through simulation, such as crystal size, decomposition temperature, and yield.

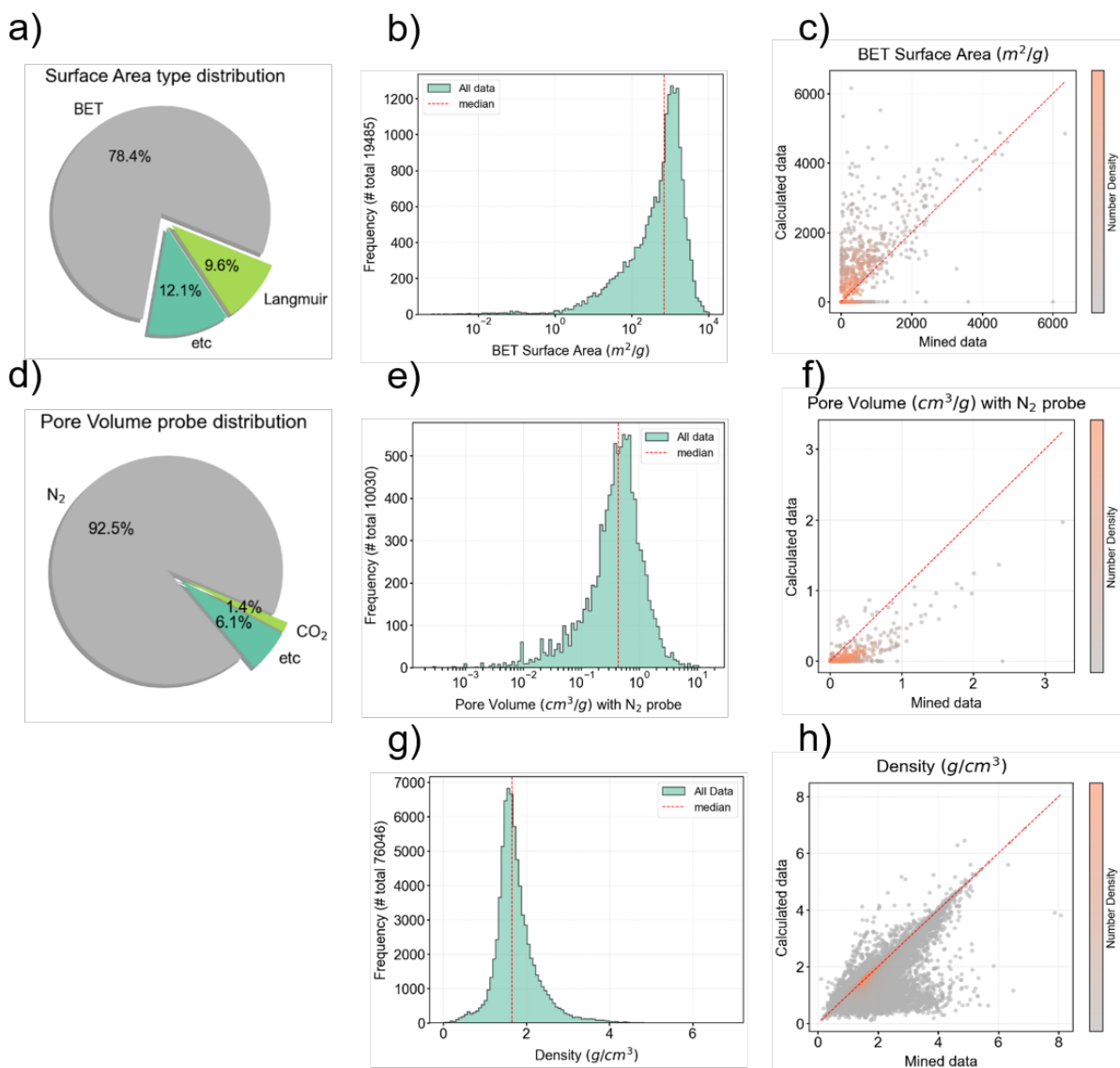

Figure 3. (a~c) Distribution of mined Surface Area data and its comparison with simulation
(d~e) Distribution of mined Pore Volume data and its comparison with simulation
(g~h) Distribution of mined Density data and its comparison with simulation

Figure 3 provides a more detailed distribution of properties such as pore volume and surface area among various properties. More detailed distributions of other properties can be found in Supplementary Figure S4. As seen in Figure 3(a), the most widely used methodology for measuring the surface area of MOF materials is the BET method, and a comparison between the mined BET surface values and the calculated BET surface area values based on structure can be found in Figure 3(c).

Similarly, as shown in Figure 3(d), $N_2$ is most commonly used as a probe for measuring the pore volume of MOFs, and a comparison between the extracted values from documents using $N_2$ as a probe and simulation calculated values can be found in Figure 3(f). The reason for the fewer data points in Figures 3(c) and 3(f) compared to 3(b) and 3(e) is that simulation values can only be calculated when the structure is reported along with the data values in the literature.

Examining Figures 3(c), 3(f), and Figure 3(h), it is clear that there is a discrepancy between the simulation and the experimental data. The reasons for this discrepancy are numerous. Simulations assume ideal conditions, whereas experiments are affected by various environmental factors such as temperature, pressure, and humidity, which are not fully considered in simulations. Furthermore, simulations often assume a perfectly ordered and defect-free structure, whereas real MOFs may exhibit surface roughness, defects, and irregularities. In addition, simulations often overlook the contribution of guest molecules. In addition to these factors, differences can arise due to various factors such as methodological differences in measurements and failure to account for intermolecular interactions in calculations.

As mentioned above, Figures 3(c), 3(f) and 3(h) show the discrepancy between experimental and simulation data resulting from various assumptions made in the simulation process. This discrepancy has a significant impact on the accuracy of predicting experimental values between models trained on simulation and experimental data. Therefore, to accurately understand the properties of MOFs and to efficiently develop materials based on them, it is crucial to use experimental data for prediction. To validate this, we first used a descriptor-based model[29] to see how the difference can affect to the accuracy of prediction. Next, to investigate how the prediction accuracy varies with machine learning model, we utilized three machine learning models: the descriptor-based model, CGCNN[30], and MOFTransformer[31]. Supplementary Note S3 provides relevant information about these machine learning models. For machine learning model prediction, we chose density as the target property because it can be easily obtained through data mining. The density prediction data contained 30,892 samples, which is sufficient for machine learning training.

*Machine learning result – Effect of training data type on prediction*

To investigate the impact of training data on predicting experimental outcome, we conducted an experiment using a descriptor-based model. We prepared two sets of training data; both containing the information on the same MOFs. One set was compiled from experimental data, while the other was derived from simulation data. We then trained two descriptor-based models separately: one with the experimental data and the other with the simulation data. Both models were used to predict outcomes on the same set of experimental test data. The model trained on experimental data achieved an R2 value of 0.802, indicating a strong correlation between predicted and actual values. In contrast, the model trained with simulation data showed a lower R2 value of 0.483. Parity plots graph predicted values against actual values to highlight discrepancies in model predictions. Figure 4 shows that models trained on simulation data deviate from the trend line when predicting experimental outcomes. This suggests that models trained on experimental data are more accurate in forecasting experimental values.

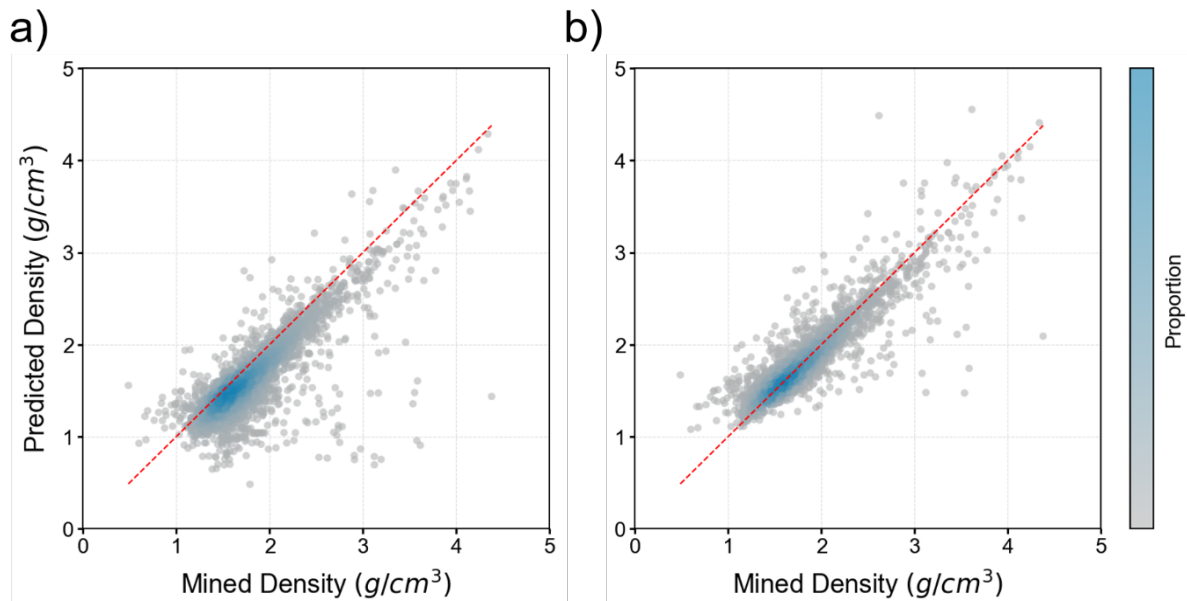

Figure 4. Parity plot of predicted value and real value when using a model that trains simulation data (a) and a model that trains experimental data (b). Model (a) was trained on simulation data, while model (b) was trained on experimental data. Both models predicted the same experimental data values during testing. The R2 values for models (a) and (b) are 0.483 and 0.802, respectively.

*Machine learning result – Effect of machine learning models on prediction*

To evaluate the accuracy of machine learning models in predicting outcomes, we trained three models using both simulation and experimental data. These models were then used to forecast experimental values, and we assessed their accuracy using five distinct sets of test and training data for each model. The results, including average accuracy and standard deviation for each dataset, are summarized in Table 2, while Table S6 provides details of the R2 values obtained in each trial. The study found that all three models were highly accurate in predicting experimental data when trained with experimental data. However, when the models trained on simulation data were applied to predict experimental outcomes, a noticeable drop in precision was observed. This difference highlights the inherent dissimilarities between experimental and simulation data, emphasizing the significance of matching the training data type with the data type being predicted for optimal accuracy.

| R2 Score | | Descriptor-based | | | | Graph-based | Multi-modal |
|---|---|---|---|---|---|---|---|
| Train | Test | RF | XGBoost | SVM | KNN | CGCNN | MOFTransformer |
| Exp | Exp | 0.793(±0.016) | 0.791(±0.016) | -0.052(±0.005) | 0.470(±0.021) | 0.815(±0.028) | 0.892(±0.010) |
| Sim | Sim | 0.762(±0.018) | 0.748(±0.016) | -0.017(±0.004) | 0.408(±0.015) | 0.930(±0.009) | 0.973(±0.016) |
| Sim | Exp | 0.469(±0.037) | 0.486(±0.035) | -0.257(±0.010) | 0.134(±0.046) | 0.400(±0.064) | 0.380(±0.060) |

Table 2. Average and standard deviation of density prediction accuracy for the descriptor-based model[32-35], CGCNN, and MOFTransformer. The three train/test datasets contain the same list of MOFs. However, the first dataset includes experimental data for both the training and test sets, the second dataset includes simulation data for both the training and test sets, and the last dataset includes simulation data for the training set and experimental data for the test set.

When applying machine learning to three different models, clear variances in performance were noted, as evidenced by the R2 values. The descriptor-based model showed the lowest performance, followed by CGCNN, and MOFTransformer showed the highest performance when the data used for training is congruent with the type of data being predicted. These differences are due to the ability of each model to accurately represent the features of MOFs when converting these features into a numerical form. The descriptor-based model relies on basic attributes, such as the atomic weight of the metals and their composition. On the other hand, CGCNN takes a nuanced approach by utilizing the relationships between atoms and bonds within graph structures. This enables it to consider details such as the type of bond, the distance between atoms, and the solid angles. However, MOFTransformer surpasses CGCNN by incorporating both local and global features derived from energy-grid embeddings, offering a more comprehensive representation than CGCNN, which primarily focuses on local characteristics. As a result, MOFTransformer achieves the highest accuracy among the tested models.

When training models on simulation data and subsequently using them to predict experimental outcomes, a reversal in trends was observed compared to when the models predict based on the type of data they were trained on. This shift is due to the distinct patterns present in experimental versus simulation data. Advanced models, which excel at capturing the intricacies of their training datasets, naturally provide more precise forecasts for data that closely resemble their training inputs. However, the effectiveness of these models may decrease when predicting outcomes for data that differ in trends from their training sets. This is because the models are specifically tuned to the characteristics of their training data.

The difference in performance is evident in the R2 values: CGCNN and MOFTransformer achieved high scores of 0.93 and 0.97, respectively, when predicting simulation data. However, their accuracy in predicting experimental data decreased, with CGCNN achieving 0.815 and MOFTransformer achieving 0.892. This suggests that while techniques such as atom-based graph embeddings or energy-grid embeddings are effective for simulation data, predicting experimental data accurately requires the consideration of additional variables.

To improve predictions of experimental data, it is essential to train machine learning models with relevant experimental datasets. Moreover, to ensure a comprehensive analysis, it is important to expand traditional methods of turning data into numerical features to encapsulate factors beyond the standard featurization processes, including environmental influences. This approach highlights the importance of not only relying on sophisticated vectorization but also integrating comprehensive datasets and modeling techniques that account for the complex realities that influence experimental results.

## Methods

*Journal Paper Retrieval & Crawling*

Our research employed a comprehensive data mining approach to collect a significant corpus of journal papers relevant to Metal-Organic Frameworks (MOFs). We targeted publications from four major scientific publishers: the American Chemical Society (ACS), Elsevier, the Royal Society of Chemistry (RSC), and Springer. We amassed a total of 41,681 unique papers, contributing to a rich dataset for our analysis. The papers were distributed among the publishers as follows: ACS contributed 16,481 papers, Elsevier 18,778, RSC 4,775, and Springer 1,647. Additional statistics on journal crawling can be found in the Supplementary Figure S5.

To maintain the integrity of our research process and respect copyright laws, we downloaded the journal papers in XML or HTML formats, adhering to ethical guidelines and obtaining explicit approval from each publisher to ensure compliance with their data usage policies.

A substantial portion of our dataset, which includes 23,091 papers, is directly linked to structures identified as part of the MOF substance family within the Cambridge Structural Database (CSD)[28], providing a robust foundation for our research. Additionally, we utilized specialized crawling code to automatically identify and retrieve 18,590 articles from Elsevier's Scopus database, leveraging the Elsevier Scopus API for efficient filtering. This process targeted articles based on their relevance to MOF substances, discerned through related terms in their titles, abstracts, and keywords, thereby ensuring a comprehensive and relevant collection of studies for our MOF research dataset.

*Integration with the CSD database*

To improve the depth and usefulness of our dataset, we integrated our collection of journal papers with the Cambridge Structural Database (CSD)[28], a well-known repository for crystallographic data. This integration was made possible by using Digital Object Identifiers (DOIs) associated with our crawled papers. By querying the CSD database with these DOIs, we determined the presence and relevance of each paper within the CSD. This ensures a seamless linkage between our dataset and the valuable crystallographic information housed in the CSD.

The CSD database contains detailed crystallography data, providing access to structure files (e.g., CIF files), refcodes, lattice information, and more. These elements are essential for our research as they provide insights into the molecular and crystalline structures of MOF substances and serve as valuable metadata for our dataset. The integration process utilized the csd-python-api version 3.0.14[36], which is compatible with the 2022 version of the CSD software. This API facilitated automated queries and retrievals from the CSD, streamlining the process of enriching our dataset with high-quality, relevant crystallographic data.

*Structure Refinement and Zeo++ Calculation*

To prepare the structures sourced from the CSD database for analysis, we standardized their representation by adopting P1 symmetry for all structures to ensure consistency across the dataset. We removed floating solvents from the structures during the refinement process to eliminate potential interference. However, we retained binding ligands that are crucial for the structural and functional integrity of the Metal-Organic Frameworks (MOFs).

In addition, structures obtained from the CSD database may present challenges such as disorders or atom duplications, which can complicate analysis. To ensure the quality of our dataset, we excluded structures with atoms located too closely together, specifically those with interatomic distances less than 1 Angstrom. This selection criterion helps to avoid inaccuracies in the dataset due to overlapping or unrealistic atomic arrangements. After the refinement stage, properties such as the density, pore volume and surface area of the structures were calculated using zeo++ software[37] with 1.82 probe radius. This tool is widely recognized for its capability in analyzing the geometric properties of nano-porous materials.

*Prompt Engineering and Find-tuning*

The data extraction process began by analyzing individual paragraphs within each journal paper. To prepare the text for processing, we utilized the Python libraries beautifulsoup4[38] and chemdataextractor[39]. These tools were effective in separating the papers into distinct paragraphs and eliminating any unnecessary HTML and XML formatting, resulting in a clean dataset for further analysis.

To orchestrate interactions between different large language models (LLMs), we used the Langchain Python library[40], specifically version 0.0.268. This setup allowed for a fully automated data extraction workflow, which was divided into three critical stages: categorization, property inclusion, and extraction. The Langchain library was instrumental in chaining these LLMs and automating the entire process. The automation was further streamlined by additional refinements, such as monitoring token length with the tiktoken Python library[41] and ensuring output consistency in TypeScript format.

For the categorization and property inclusion stages, we employed LLMs fine-tuned on OpenAI's GPT-3.5-turbo model, tailored with 723 and 681 example sets, respectively. This fine-tuning was crucial for achieving high accuracy in classifying and identifying relevant data, with examples developed through GPT-4[42] prompt engineering and refined with human review. Figure 5(a) illustrates the categorization process, employing finely tuned, concise prompts and examples for precise text category identification. Figure 5(b) depicts the inclusion step by the synthesis agent, adopting a similar approach to categorization but with explicit rules and human-provided examples to assure task accuracy, especially in distinguishing various synthesis conditions.

The final extraction step shifted to prompt engineering with GPT-4, chosen for its adaptability in processing novel information, with occasional support from GPT-3.5-turbo-0125 for tasks exceeding GPT-4's 8000-token limit. The introduction of the GPT-4-32K API was not incorporated due to the project's advanced stage and budget considerations. Figure 5(c) explores the extraction methodology for the synthesis condition agent, employing a detailed prompt strategy that includes a comprehensive rule set and specifies the data format, notably using JSON for consistency.

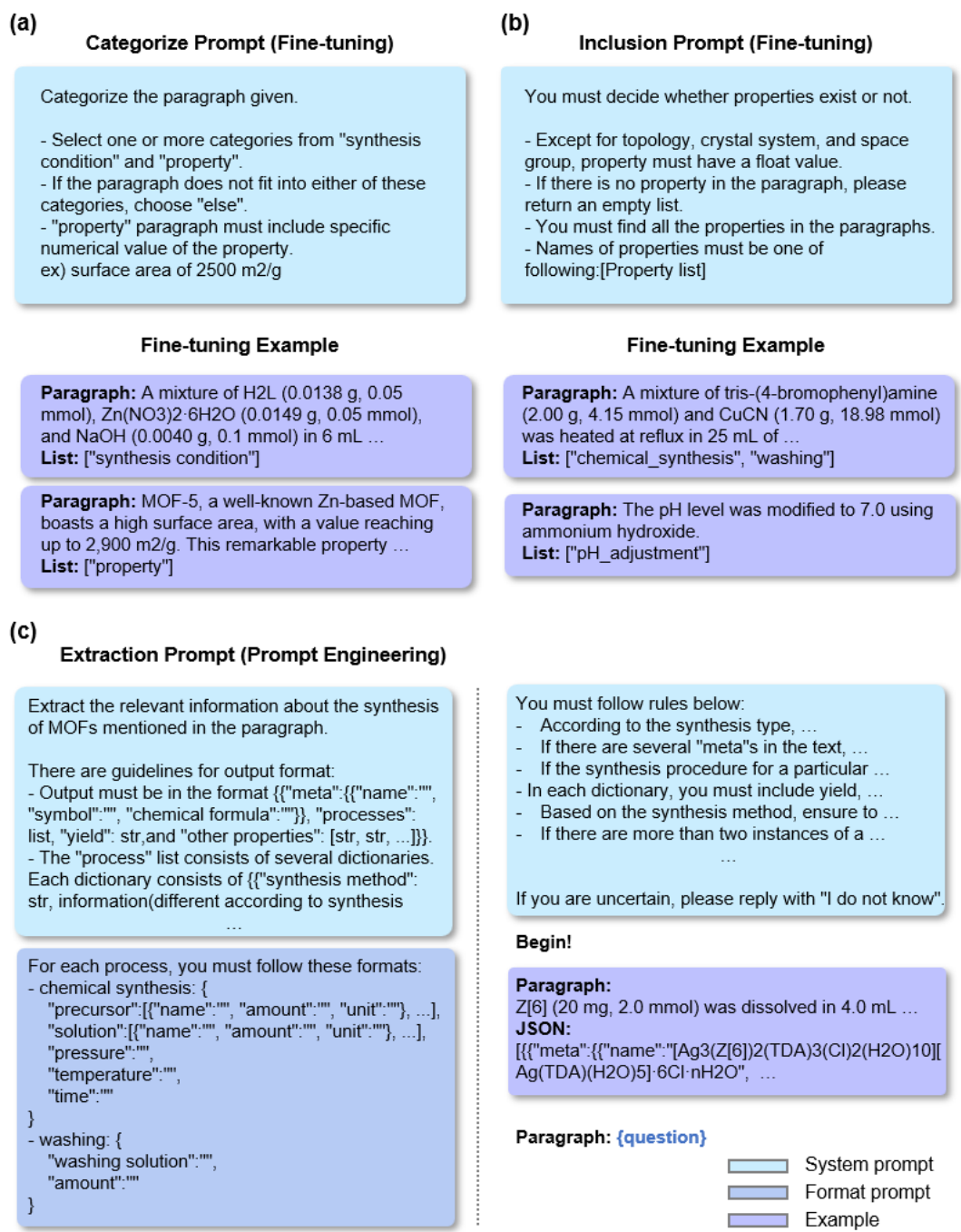

Figure 5. (a) An example prompt for categorization task (b) An example prompt for inclusion task (c) An example prompt for extraction prompt

**ChatMOF Integration (Optional)**

With the development of LLMs, there has been a significant increase in interest in their accuracy and applicability in various fields[43-51], including materials science. Researchers in this field are particularly interested in using chatbots to obtain accurate synthetic information and to answer unstructured questions. Yaghi et al.[27] proposed an innovative approach to addressing these challenges by introducing a chatbot designed to navigate and interpret synthetic datasets. Through their work, they envisioned transforming complex datasets into a dynamic, interactive dialogue system, making the data more accessible and understandable to users. This initiative led to the creation of the ChatGPT Chemistry Assistant, designed as a reliable and knowledgeable companion for exploring chemical reactions, with a particular focus on MOF synthesis.

Recent research has focused on extending the capabilities of LLMs by integrating them with a variety of tools[51,52]. Innovations such as babyAGI[53] and AutoGPT[54] are at the forefront of these efforts, demonstrating the potential of LLMs when combined with external tools to perform tasks that were previously beyond their scope. A recent development in this area is ChatMOF[55], a chatbot tailored for the prediction of properties and inverse design of MOFs. ChatMOF offers a variety of services related to MOFs, such as database searching, property prediction using MOFTransformer[31,56] with universal transfer learning, and the generation of materials with user-desired properties using genetic algorithms.

Figure 6 and Supplementary Figure S10 demonstrate the functionality of this chatbot by linking ChatMOF with an L2M3 database. Supplementary Figure S10 introduces two new tools: the Search Synthesis Condition tool, which finds synthesis details and related articles using user questions, and the Material Finder, which translates common MOF names (such as IRMOF-1 and MIL-102) to their CCDC database REFCODEs. Figure 6 explains how ChatMOF finds the synthesis details of MIL-120. It uses Material Finder to match MIL-120 to its REFCODE and then employs the Synthesis Condition tool to retrieve the synthesis information. This enables ChatMOF to provide the requested synthesis details. Additionally, Supplementary Figure S11 demonstrates how ChatMOF obtains the metadata for the MOF-649 synthesis article, indicating its compatibility with the L2M3 database.

The integration of ChatMOF into an autonomous artificial intelligence system represents a significant leap into the future of materials science research. While traditional LLMs have been limited to processing text-based data, the envisioned future expands their capabilities to provide accurate and quality-assured information in the field of materials science using a variety of specialized tools. This evolution marks a significant advance in the utility and applicability of LLMs in scientific research, particularly in the field of materials science.

> **Give me a numbered step-by-step explanation of the synthesis information in MIL-120.**

**Thought**: I need to change MIL-120 to refcode
**Action**: material_finder
**Action Input**: Convert the name MIL-120 to refcode

*(…run material finder tool)*

**Observation**: The refcode for MIL-120 is BUSQIQ.
**Thought**: I need to find the synthesis information for BUSQIQ.
**Action**: search_synthesis_condition
**Action Input**: "Search for the synthesis conditions of BUSQIQ"

*(…run search_synthesis_condition tool)*

**Observation**: The synthesis conditions of BUSQIQ involve several steps. The first step …
**Thought**: I now know the final answer

**Final Answer:**
The synthesis of MIL-120 (refcode: BUSQIQ) involves the following steps:

1. Hydrothermal synthesis using precursors aluminum nitrate (3.2 g), pyromellitic acid (0.5 g), and sodium hydroxide solution (3.4 mL). The solvent used is deionized water (20 mL). The synthesis is carried out at a temperature of 210°C for 24 hours.
2. pH adjustment to 12.2.
3. Filtration.
4. Washing with 300 mL of deionized water.
5. Drying at room temperature.

Figure 6. ChatMOF combined with L2M3 database. Casestudy of MIL-120

*Conference.* 804-823 (PMLR).

51  Bran, A. M., Cox, S., White, A. D. & Schwaller, P. Chemcrow: Augmenting large-language models with chemistry tools. *arXiv preprint arXiv:2304.05376* (2023).

52  Shen, Y. *et al.* Hugginggpt: Solving ai tasks with chatgpt and its friends in hugging face. *Advances in Neural Information Processing Systems* **36** (2024).

53  yoheinakajima. *babyagi*, <https://github.com/yoheinakajima/babyagi> (

54  Significant-Gravitas. *AutoGPT*, <https://github.com/Significant-Gravitas/AutoGPT> (

55  Kang, Y. & Kim, J. Chatmof: An autonomous ai system for predicting and generating metal-organic frameworks. *arXiv preprint arXiv:2308.01423* (2023).

56  Park, H., Kang, Y. & Kim, J. Enhancing Structure–Property Relationships in Porous Materials through Transfer Learning and Cross-Material Few-Shot Learning. *ACS Applied Materials & Interfaces* **15**, 56375-56385 (2023).28

Supplementary information for:

# L2M3: Large Language Model based Material Miner

Wonseok Lee[1†], Yeonghun Kang[1†], Taeun Bae[1†], Jihan Kim[1*]

[1]Department of Chemical and Biomolecular Engineering, Korea Advanced Institute of Science and Technology, Daejeon, Republic of Korea

\* Correspondence to: Jihan Kim (jihankim@kaist.ac.kr)

[†] *Wonseok Lee, Yeonghun Kang and Taeun Bae contributed equally to this paper.*

# Table of Contents

**Supplementary Notes S1. Explanation of the evaluation process and criteria**

For the evaluation of MOF synthesis papers published by ACS, Elsevier, RSC, and Springer, 25 papers each were randomly selected. Additionally, 25 papers each were randomly chosen from MOF application papers published by ACS and Elsevier. In total, 1069 paragraphs from the 150 selected papers underwent evaluation.

During the assessment of the categorization task, we considered three evaluation criteria. The first criterion is the true case, as shown in Supplementary Figure S6 a, where the agent accurately identifies the property or synthesis condition within the paragraph. In the table, this refers to correctly identifying which category - crystal information, property information, or bond & angle information - the table belongs to. Secondly, there are false positive cases, as depicted in Supplementary Figure S6 b, where a property or synthesis condition is believed to exist, but in reality, it either does not exist at all or contains different information. Lastly, there are false negative cases, as shown in Supplementary Figure S6 c, where the agent determines that there is no information, but in fact, information does exist.

The Inclusion task is divided into three evaluation criteria. The first criterion is the true case, as illustrated in Supplementary Figure S7 a, where all information within the paragraph is accurately identified. In this scenario, all properties or synthesis processes within the paragraph are listed, and each component is individually assessed. As shown in Supplementary Figure S7 a, correctly extracting all three synthesis process pieces of information would be marked as three correct answers. In the example, only one paragraph is evaluated. However, in real, all information from multiple paragraphs of the same type is extracted simultaneously during Inclusion task. Additionally, there are two cases to consider: the false negative case (depicted in Supplementary Figure S7 b), where information within the paragraph is either not listed or listed with incorrect names, and the false positive case (shown in Supplementary Figure S7 c), where information not present within the paragraph is falsely listed.

  The Extraction task applies similar scoring criteria to the preceding tasks. Accurately extracted information is considered a true case, as shown in Supplementary Figure S8 a. Scoring is based on individual keys within each dictionary. If all components associated with a key are correctly extracted, it is marked as correct. For example, in the case of 'meta', correct extraction of name, symbol, and chemical formula information results in a correct mark. Similarly, for 'magnetic susceptibility', correctness is determined by the accurate extraction of the value, unit, temperature, and condition. Synthesis conditions are evaluated separately for each method. In solvothermal synthesis, for instance, details such as precursor, solvent, surfactant, pressure, and temperature are assessed individually. All extracted information is considered correct only if all details are accurately extracted. For example, when dealing with the term 'precursor', each instance within a paragraph is evaluated separately for its name, amount, and unit. If all details for each precursor are accurately extracted, it is marked as correct. However, if any information within a paragraph is not listed correctly, as demonstrated in Supplementary Figure S8 b, it is classified as a false negative case. If information that does not exist is fabricated and listed, as shown in Supplementary Figure S8 c, it is classified as a false positive case. Similar to the Inclusion task, in the Extraction task, information is extracted from multiple paragraphs simultaneously during the actual assessment process. All information within these paragraphs is extracted collectively.

**Supplementary Notes S2. Comparison of performance with previous studies**

Several papers have conducted text mining on MOF papers. Most of them focused on extracting specific information, such as a certain property or synthesis condition of MOF. To ensure a fair comparison, we conducted accuracy measurements based on the types of data extracted from different papers and compared the results with each respective paper.

The study compared its performance with research focused on extracting properties. Park et al.[1] reported an accuracy of 73.2% for surface area and 85.1% for pore volume when mining on approximately 200 sets of surface area and pore volume. In this study, there were approximately 100 items assessed for pore volume and about 150 for surface area. The evaluation results showed an accuracy of 100% for pore volume and 99.4% for surface area. Nandy et al.[2] extracted information on approximately 3000 decomposition temperatures, while our study extracted decomposition temperatures for over 20,000 items. Tayfuroglu et al.[3] extracted 6000 surface area and 7500 pore volume data from nearly 60,000 papers. However, when we randomly sampled 100 data points from their extracted dataset and assessed accuracy, we observed relatively low accuracy ranging from 70% to 80%.

Next, we compared our study with papers that focus on extracting synthesis conditions. In Luo et al.'s study[4], they categorized synthesis paragraphs and extracted information on time, temperature, solvent, and additive. They performed extraction on approximately 6,000 papers and demonstrated high accuracy of over 90% in both tasks. In our study, we performed data mining on over 40,000 papers and achieved high F1 scores close to 0.9 for both tasks. Park et al.[5] reported that they conducted text mining on approximately 30,000 papers. For both the synthesis paragraph categorization task and the extraction task of MOF, precursor, and solvent, the F1 scores were close to 0.9. In our study, the F1 score for each task was higher than 0.95. In Glasby et al.'s study[6], mining was conducted on over 40,000 papers to measure accuracy for synthesis routes, topologies, linkers, and metal precursors. The results showed a relatively lower accuracy with an F1 score of around 0.5. In our research, we confirmed achieving high accuracy of around 0.9. Zheng et al.[7] distinguished three processes to measure individual performances in their work. Process 1 involved extracting desired information from given synthesis paragraphs and demonstrated a very high F1 score of 0.96. However, this was observed for only about 200 papers with specific formats, indicating that the high F1 score might be limited to such papers. For Process 2, which is similar to the classification task in our research, our study achieved a higher F1 score.

**Supplementary Note S3. Training details for three machine learning models**

The descriptor-based model[8] uses different MOF features to generate a single vector, which serves as input for a machine learning model. The model employs three main types of descriptors: metal features, linker features, and global features. Metal features comprise six attributes, namely atomic number, atomic weight, atomic radius, Mulliken electronegativity, polarizability, and electron affinity of the metal. The Linker features consist of 210 RDKit descriptors[9] and 1024 Morgan fingerprints[10], which are extracted from the SMILES notation[11] of the linker. Global features include 120 Meredig descriptors[12] that encompass elemental fractions, average atomic mass, and other relevant properties. The number of vectors for metal and linker features is determined based on the maximum number of metals and linkers per MOF, as shown in Figure S9. The machine learning models used in this study were selected by comparing their performance using the Python PyCaret module[13]. The top-performing models, including Random Forest[14], XGBoost[15], SVM[16], and KNN[17], were chosen based on this comparison. The hyperparameters for each model were optimized using GridsearchCV[18]. For Random Forest, the n_estimators were set to 100, and for XGBRegressor, the n_estimators were also set to 100. SVM used rfb as the kernel, and KNN set n_neighbors to 5.

The CGCNN[19], or Crystal Graph Convolutional Neural Network, is a model specifically designed for the periodic crystal system. It represents atoms and bonds of a crystal as nodes and edges in a graph, respectively. Using a convolutional neural network, it learns to consider the relationships with neighboring atoms, achieving higher accuracy. The architecture of CGCNN consists of 5 convolution layers, followed by 1 hidden layer after pooling. The convolution layers have 64 hidden atom features, and there are 128 hidden features after pooling. The models are trained using the adam optimizer with a learning rate of 0.01 and a batch size of 16. The training process lasts for 40 epochs. For more information, please refer to the original CGCNN paper.

In the case of MOFTransformer[20], the AdamW optimizer was used during the pre-training step with a learning rate set to $10^{-4}$ and weight decay set to $10^{-2}$ for all three tasks. The model was trained with a batch size of 1024 for 100 epochs. For the pre-training dataset, it was randomly divided into training, validation, and test sets with sizes of 800,000, 126,611, and 100,000 respectively. During fine-tuning, the model was trained with a batch size of 32 for 40 epochs. The optimizer and learning rates remained the same as those used during the pre-training step. The fine-tuning dataset was split into training, validation, and test sets in an 8:1:1 ratio.

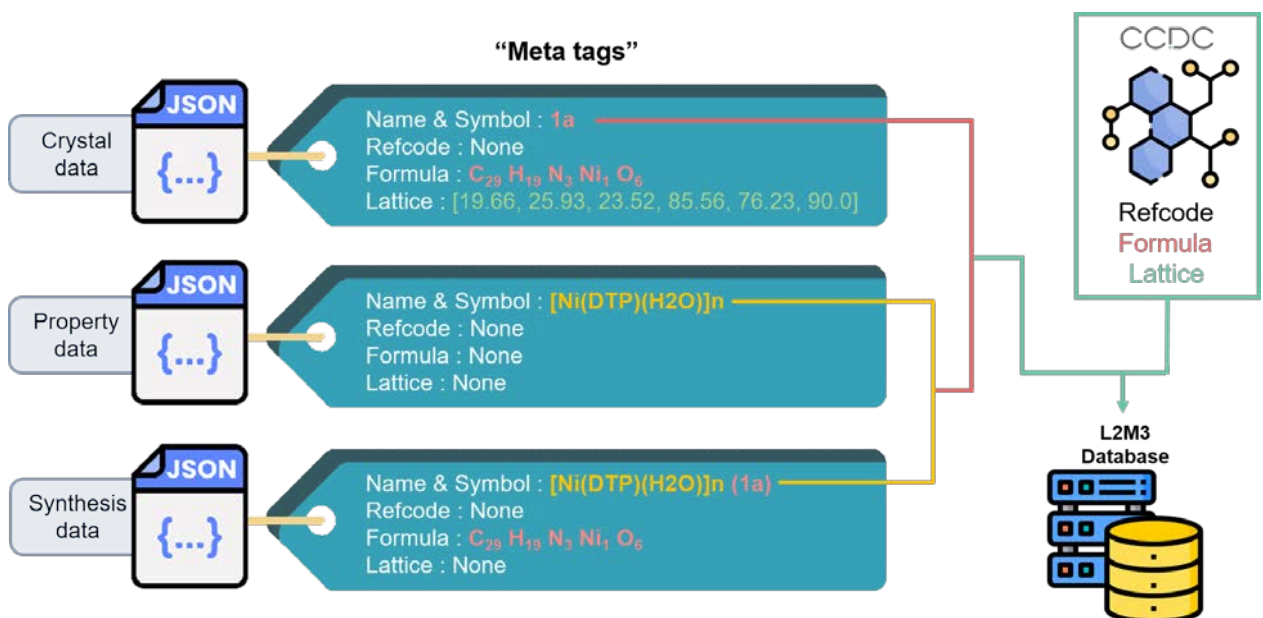

**Supplementary Figure S1.** Schematic figure for a matching agent. The purpose of the matching agent is the standardization of material names and symbols. Combine the name, symbol, and structural data to create a single dataset for each material.

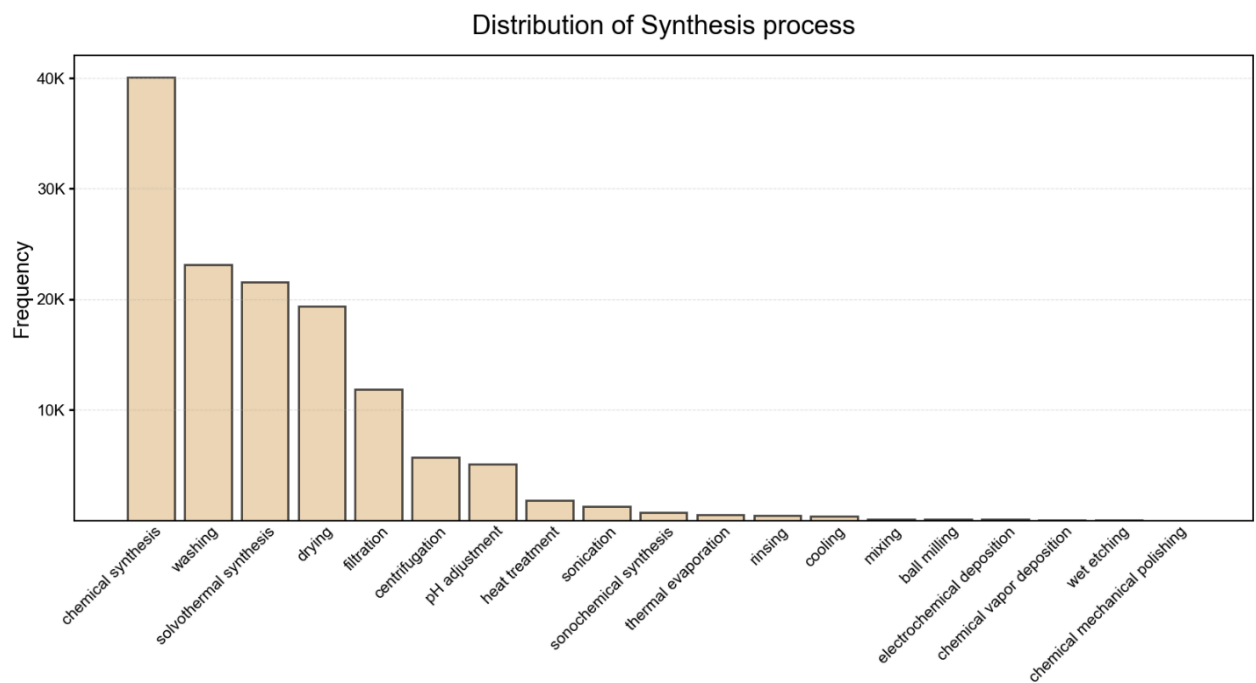

**Supplementary Figure S2. Figure for distribution of synthesis process**

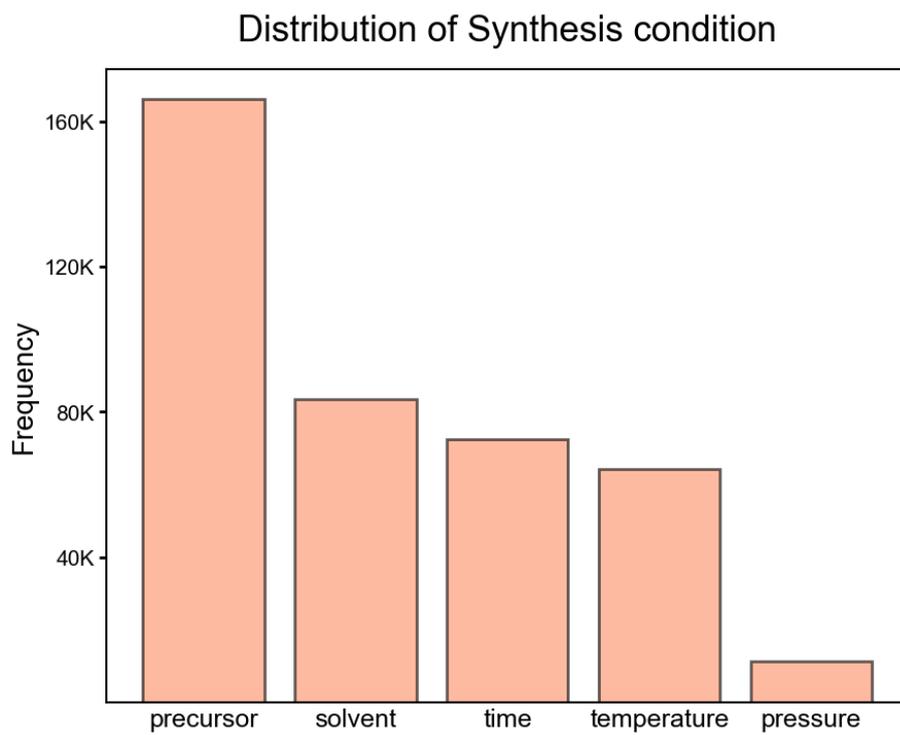

**Supplementary Figure S3. Figure for distribution of synthesis conditions**

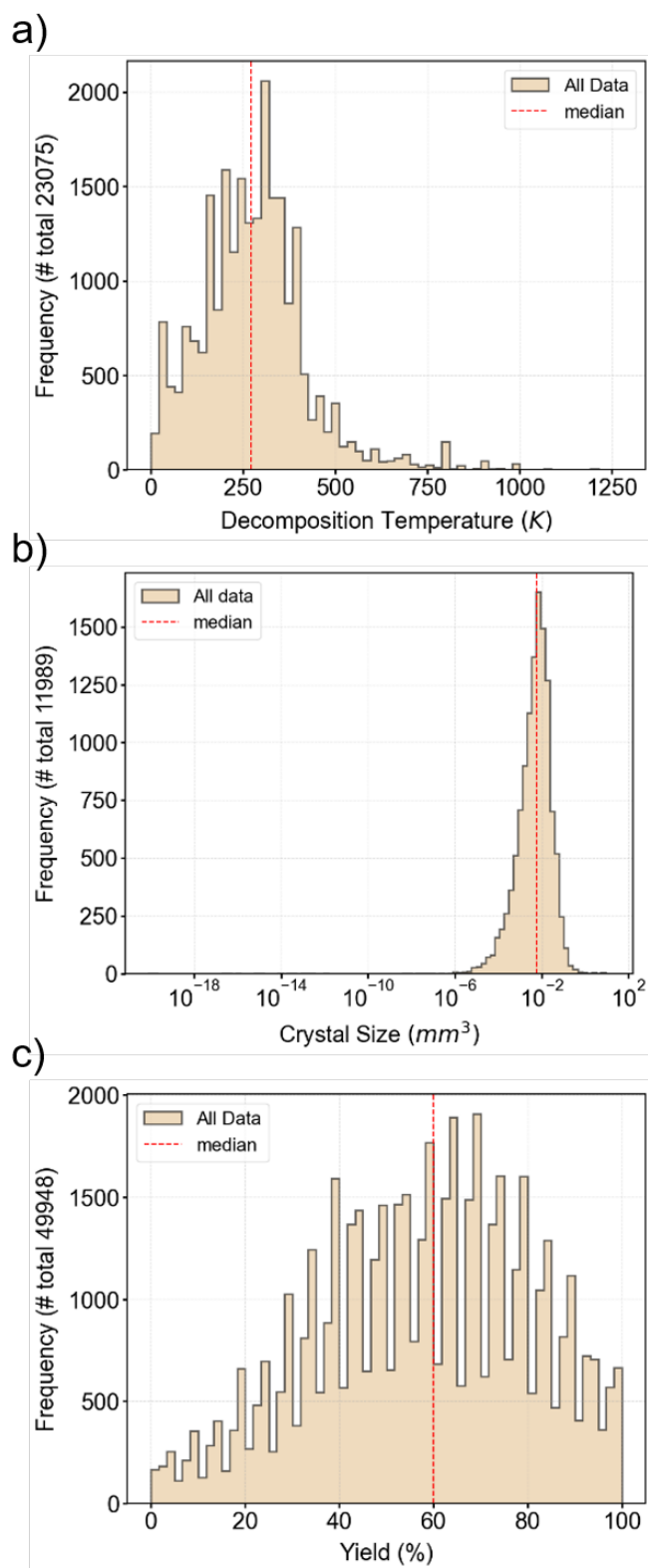

**Supplementary Figure S4. Distribution of extracted properties** Distribution of a. Decomposition Temperature, b. Crystal Size, c. Yield

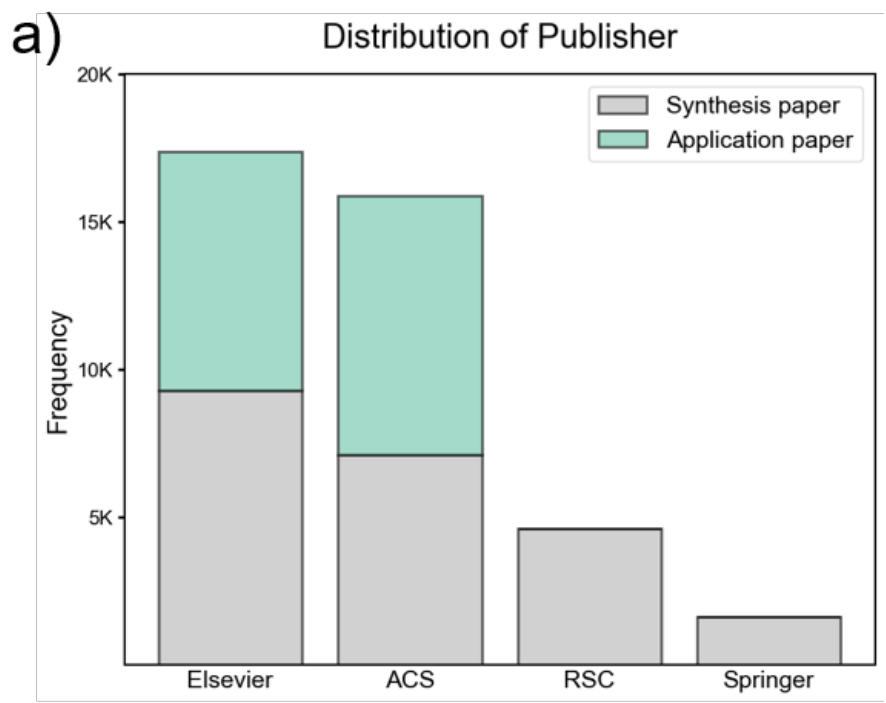

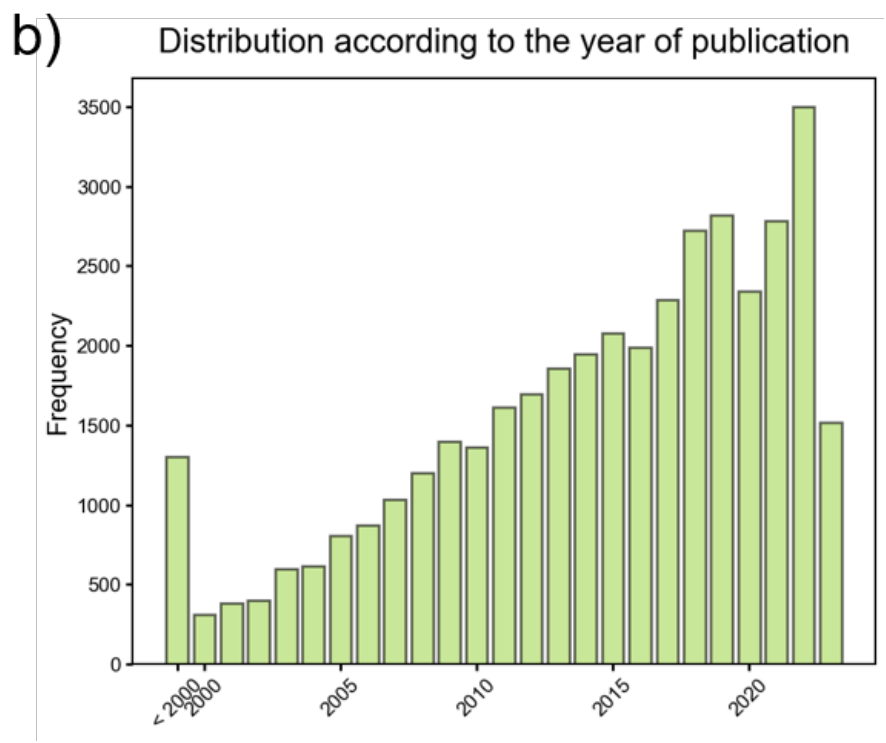

**Supplementary Figure S5. Distribution of crawled paper** a. Distribution based on the publisher b. Distribution according to the year of publication

**Supplementary Figure S6**. a. Examples of true cases b. An example of false positive case that misunderstands 'property' as 'synthesis condition' c. An example of false negative case that misunderstands 'property as 'else

**Supplementary Figure S7.** a. Examples of right cases b. An example of false negative case that misunderstands 'solvothermal_synthesis' as 'chemical_synthesis' c. An example of false positive case that hallucinates 'porev_volume' even though 'pore_volume' information does not exist

**Supplementary Figure S8** a. Correct example of property extraction and synthesis condition extraction b. An example of false negative case that the agent does not extract existing information c. An example of false positive case that the agent hallucinates the information

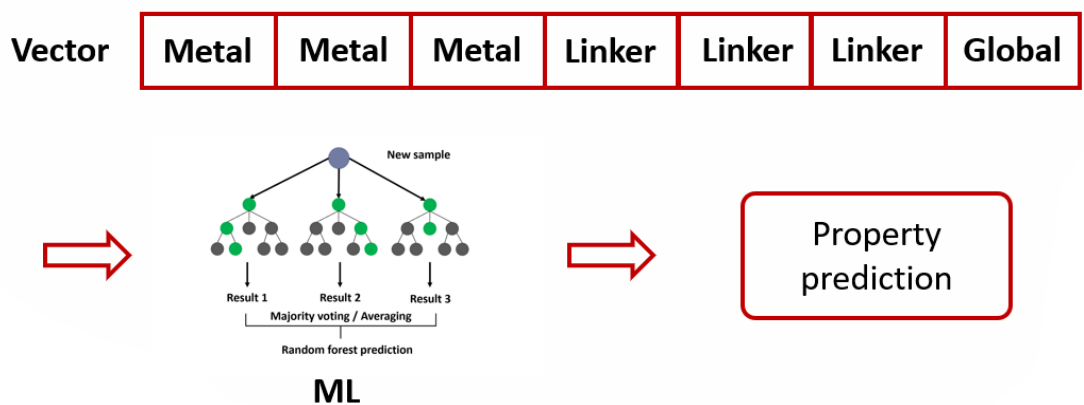

**Supplementary Figure S9.** An overview of descriptor model

**[Agent]** Planning, selecting appropriate tools, and interpreting the results using GPTs.

**[Tool] Search synthesis condition**
Thought: To find the synthesis paper for a …
Input :
```
print (df.loc[df['name'] == 'BOQQEF',
    ['doi', 'title', 'date', 'journal', 'author_list']
    ].to_markdown())
```
Observation:

| | doi | title | … |
|---|---|---|---|
| 1 | 10.1039/.. | Synthesis and hy… | … |

**[Tool] Material Finder**
Thought : The question is asking to convert …
Action: convert to refcode
Action Input : MIL-120
 *(Find the similar MOF-name)*
Observation:

| BUSQIQ | ZIBGID | WEVYEE | … |
|---|---|---|---|
| MIL-120 | MIL-102 | MIL-97(Al)-HR | … |

Output : The refcode for MIL-120 is BUSQIQ.

**Supplementary Figure S10. Two tools of chatbot system**

**User:** Can you give me the metadata of the synthesis paper of MOF-649?

**Agent:**
Thought: I need to change MOF-649 to refcode
Action: material_finder
Action Input: Convert the name MOF-649 to refcode

(…run material_finder tool)

Observation: The refcode for MOF-649 is BOQQEF.
Thought: I need to find the synthesis paper for BOQQEF
Action: search_synthesis_condition
Action Input: Search for the synthesis paper for BOQQEF

(…run search_synthesis_condition tool)

Observation: The synthesis paper for BOQQEF is titled "Synthesis and hydrogen adsorption …
Thought: I now know the final answer

**Final Answer:**
The synthesis paper for MOF-649 (BOQQEF) is titled "Synthesis and hydrogen adsorption properties of internally polarized 2,6-azulenedicarboxylate based metal–organic frameworks". It was published in 2014 in the Journal of Materials Chemistry A. The DOI is 10.1039/C4TA04393F. The authors are Samir Barman, Anupam Khutia, Ralph Koitz, Olivier Blacque, Hiroyasu Furukawa, Marcella Iannuzzi, Omar M. Yaghi, Christoph Janiak, Jürg Hutter, and Heinz Berke.

**Supplementary Figure S11. Figure for chatbot casestudy: MOF-649**

**Table 3.** Three-component synthesis of propargylamines using pyrrolidine catalyzed by the Cu(I)-MOF.a

| Entry | R$^1$ | R$^2$ | Time (h) | Yield (%)$^b$ |
|---|---|---|---|---|
| 1 | H | Ph | 24 | 89 |
| 2$^c$ | H | 4-MeOC$_6$H$_4$ | 26 | 58 |
| 3 | CH$_3$ | Ph | 26 | 86 |
| 4 | Cl | Ph | 24 | 62 |
| 5 | MeO | Ph | 24 | 59 |

a. Unless otherwise indicated, all reactions were carried out with 1b (0.50 mmol), 2 (0.50 mmol), and 3 (0.50 mmol) and the MOF catalyst (0.0125 mmol, 2.5 mol%) under neat conditions at 80 °C
b. Yield of the isolated product after column chromatography.
c. With a catalyst loading of 10 mol %

**Supplementary Table S1.** Example of a property table.
(Source: Chinese Chemical Letters Volume 26, Issue 1, January 2015, Pages 6-10)

Table 2. Selected bond lengths (A), angles (deg), and dihedral angles (deg) for MOF 3 with estimated standard deviations in parentheses.

| Bond/Angle/Dihedral | Value (Å or deg) | Bond/Angle/Dihedral | Value (Å or deg) |
| --- | --- | --- | --- |
| Zn1–O1 | 1.988(2) | Zn1–O2#1 | 2.019(2) |
| Zn1–O3#2 | 2.028(2) | Zn1–N12#3 | 2.174(2) |
| Zn1–N1 | 2.199(2) | Zn1–O4#2 | 2.483(3) |
| N1–C2 | 1.336(3) | C2–C3 | 1.386(4) |
| C3–C4 | 1.396(4) | C4–C7 | 1.435(3) |
| C7–C8 | 1.195(4) | O1–C1 | 1.261(3) |
| O2–C1 | 1.251(3) | C8–O3 | 1.265(4) |
| C8–O4 | 1.236(4) | - | - |
| O1–Zn1–O2#1 | 120.15(8) | O1–Zn1–O3#2 | 147.42(9) |
| O2#1–Zn1–O3#2 | 92.41(9) | O1–Zn1–N12#3 | 89.54(8) |
| O2#1–Zn1–N12#3 | 86.20(7) | O1–Zn1–N1 | 92.36(8) |
| O3#2–Zn1–N1 | 88.54(8) | N12#3–Zn1–N1 | 176.11(8) |
| O2#1–Zn1–O4#2 | 149.35(8) | - | - |
| O1–Zn1–N1–C2 | 16.9(2) | O2#1–Zn1–N1–C2 | 137.1(2) |
| N12#3–Zn1–N1–C2 | 136.1(11) | O4#2–Zn1–N1–C2 | -73.6(2) |
| C8#2–Zn1–N1–C2 | -101.7(2) | O2#1–Zn1–N1–C6 | -40.84(19) |
| O4#2–Zn1–N1–C6 | 108.54(19) | C5–C4–C7–C8 | -145(4) |
| O2#1–Zn1–O1–C1 | -34.7(2) | O3#2–Zn1–O1–C1 | 147.51(19) |

**Supplementary Table S2.** Example of a bond table.
(Source: Z. Naturforsch. 2012, 67b, 103 – 106; received January 12, 2012)

**Table 1. Crystal structure data for MOF 3.**

| Parameter | Value |
|---|---|
| Formula | C20H12N2ZnO4 · C3H7NO |
| Molar mass (Mr) | 482.78 |
| Crystal size (mm^3) | 0.24 × 0.12 × 0.06 |
| Temperature (T, K) | 123(2) |
| Crystal system | monoclinic |
| Space group | P2_1/c (no. 14) |
| a (Å) | 10.400(1) |
| b (Å) | 19.294(2) |
| c (Å) | 10.706(1) |
| β (degrees) | 96.94(19) |
| Volume (V, Å^3) | 2132.5(4) |
| Z | 4 |
| Calculated density (D_calcd, g cm^−3) | 1.50 |
| μ (MoKα, cm^−1) | 1.2 |
| F(000) | 992 |
| hkl range | ±13, ±25, ±13 |
| 2θ_max (degrees) | 55 |
| Reflections measured / unique / R_int | 28598 / 4884 / 0.059 |
| Parameters refined / restraints / data | 284 / 61 / 4884 |
| R(F) (I ≥ 2σ(I)) / wR (F^2) (all reflections) | 0.042 / 0.089 |
| Goodness-of-fit (GoF) on F^2 | 1.05 |
| Δρ_fin (max / min), e Å^−3 | 0.56 / −0.37 |

**Supplementary Table S3.** Example of a crystal information table.
(Source: Z. Naturforsch. 2012, 67b, 103 – 106; received January 12, 2012)

| Property Name | Argument | Property Name | Argument | Property Name | Argument |
|---|---|---|---|---|---|
| Surface area | Type, Probe, Value, Unit, Condition | Heat capacity | Value, Unit, Condition | Catalytic activity | Value, Unit, Time, Condition |
| Pore volume | Probe, Value, Unit, Condition | Thermal expansion coefficient | Value, Unit, Condition | Density | Value, Unit, Condition |
| Crystal size | Value, Unit, Condition | Thermal conductivity coefficient | Value, Unit, Condition | Magnetic moment | Value, Unit, Temperature, Condition |
| Gas adsorption | Adsorbate, Adsorbed amount, Unit, Temperature, Pressure, Condition | Elastic constant | Value, Unit, Condition, Type | Magnetic susceptibility | Value, Unit, Temperature, Condition |
| Porosity | Probe, Value, Unit, Condition | Formation energy | Value, Unit, Condition | Peak spectrum | Value, Unit, Type, Condition |
| Pore diameter | Value, Unit, Condition | Adsorption energy | Gas type, Value, Unit, Condition | Cell volume | Value, Unit, Condition |
| Crystal system | Value, Condition | Henry coefficient | Gas type, Value, Unit, Condition | Lattice parameters | Value(a, b, c, alpha, beta, gamma), Condition |
| Space group | Value, Condition | Selectivity | Value, Unit, Substrate, Catalyst, Pressure, Temperature, Solvent, Time, Condition | Topology | Value, Condition |
| Chemical formula weight | Value, Unit, Condition | Conversion | Value, Unit, Substrate, Catalyst, Pressure, Temperature, Solvent, Time | Material Color | Value, Condition |
| Decomposition temperature | Value, Unit, Type, Condition | Reaction yield | Value, Unit, Substrate, Catalyst, Pressure, Temperature, Solvent, Time | Material Shape | Value, Condition |
| Simulation parameters | Symbol, Value, Unit, Type | Proton conductivity | Value, Unit, Temperature, RH, Ea, Guest | - | - |

**Supplementary Table S4.** List of properties

| Process Name | Argument | Process Name | Argument |
|---|---|---|---|
| Ball milling | Material, Type, Rotation speed, Total time, Milling time, Rest time, Ball to material weight ratio, Material mass, Solvent, Ball material, Ball size, Jar shape, Jar volume, Jar material, Atmosphere | Solvothermal synthesis | Precursor(name, amount, unit), Solvent(name, amount, unit), Reducing Agent(name, amount, unit), Surfactant(name, amount, unit), Pressure, Temperature, Time, Heating rate, Cooling rate |
| Centrifugation | Revolution per time, Relative centrifugal force, Time, Temperature, Additive(name, amount, unit), Tube material, Tube volume | Sol gel synthesis | Temperature, Precursor, Solvent, Time |
| Chemical mechanical polishing | Slurry concentration, Slurry density, Viscosity, Polishing rate, Surface roughness, Precursor(name, amount, unit), Pressure, Temperature, Atmosphere, Rpm | Sonication | Composition, Power, Solvent, Temperature, Time, Type |
| Chemical synthesis | Precursor(name, amount, unit), Solution(name, amount, unit), Pressure, Temperature, Time | Sonochemical synthesis | Precursor(name, amount, unit), Solvent(name, amount, unit), Temperature, Ultrasonic frequency, Power, Time |
| Chemical vapor deposition | Working pressure, Carrier gas name, Carrier gas flow rate, Precursor(name, amount, unit) | Thermal evaporation | Precursor(name, amount, unit), Working pressure, Substrate temperature, Deposition rate, Time |
| Drying | Pressure, Temperature, Atmosphere, Time | Wet etching | Temperature, Etchant, Concentration, Solvent, Additive, Time, Stirring rate |
| Electrochemical deposition | Electrolyte precursor, Electrolyte solvent, Electrolyte concentration, Electrolyte pH, Additive, Working electrode, Counter electrode, Mode, Voltage, Current density, Stirring, Temperature, Atmosphere, Time | Washing | Washing solution, Amount |
| Heat treatment | Heat treatment method, Atmosphere, Heat treatment temperature, Heat treatment time, Heating rate, Cooling rate | Cooling | Pressure, Temperature, Cooling rate, Time |
| Microwave assisted synthesis | Precursor(name, amount, unit), Solution(name, amount, unit), Temperature, Atmosphere, Time | pH adjustment | pH, Modulator |
| Mixing | Type, Material mix, Solvent(name, amount, unit), Rotation speed, Temperature, Time | Filtration | Time, Atmosphere, Pressure |
| Rinsing | Temperature, Time, Additive(name, amount, unit) | - | - |

**Supplementary Table S5.** List of synthesis condition type

| Model | Dataset | | 1 | 2 | 3 | 4 | 5 | Average | Standard deviation |
|---|---|---|---|---|---|---|---|---|---|
| | Train | Test | | | | | | | |
| RF | Exp | Exp | 0.802 | 0.793 | 0.787 | 0.815 | 0.768 | 0.793 | 0.016 |
| | Sim | Sim | 0.764 | 0.782 | 0.763 | 0.773 | 0.729 | 0.762 | 0.018 |
| | Sim | Exp | 0.483 | 0.489 | 0.422 | 0.521 | 0.430 | 0.469 | 0.037 |
| XGBoost | Exp | Exp | 0.805 | 0.793 | 0.777 | 0.811 | 0.770 | 0.791 | 0.016 |
| | Sim | Sim | 0.752 | 0.769 | 0.746 | 0.755 | 0.720 | 0.748 | 0.016 |
| | Sim | Exp | 0.500 | 0.502 | 0.445 | 0.536 | 0.447 | 0.486 | 0.035 |
| SVM | Exp | Exp | -0.056 | -0.048 | -0.047 | -0.060 | -0.047 | -0.052 | 0.005 |
| | Sim | Sim | -0.022 | -0.015 | -0.013 | -0.023 | -0.015 | -0.017 | 0.004 |
| | Sim | Exp | -0.272 | -0.249 | -0.249 | -0.265 | -0.248 | -0.257 | 0.010 |
| KNN | Exp | Exp | 0.488 | 0.482 | 0.446 | 0.492 | 0.444 | 0.470 | 0.021 |
| | Sim | Sim | 0.406 | 0.434 | 0.389 | 0.403 | 0.407 | 0.408 | 0.015 |
| | Sim | Exp | 0.148 | 0.183 | 0.069 | 0.177 | 0.093 | 0.134 | 0.046 |
| CGCNN | Exp | Exp | 0.824 | 0.832 | 0.794 | 0.853 | 0.773 | 0.815 | 0.028 |
| | Sim | Sim | 0.921 | 0.947 | 0.930 | 0.921 | 0.929 | 0.930 | 0.009 |
| | Sim | Exp | 0.441 | 0.464 | 0.362 | 0.441 | 0.293 | 0.400 | 0.064 |
| MOFTransformer | Exp | Exp | 0.889 | 0.893 | 0.883 | 0.910 | 0.884 | 0.892 | 0.010 |
| | Sim | Sim | 0.979 | 0.942 | 0.982 | 0.983 | 0.981 | 0.973 | 0.016 |
| | Sim | Exp | 0.413 | 0.427 | 0.311 | 0.445 | 0.304 | 0.380 | 0.060 |

**Supplementary Table S6. R2 scores for five different train/test dataset (Exp: Experiment, Sim: Simulation)**